\documentclass[prb,twocolumn,showpacs,superscriptaddress,floatfix]{revtex4}

\usepackage{amsmath,amsfonts,amssymb,bm}
\usepackage{graphicx}
\usepackage[all]{xy}

\newcommand{\proj}[2]{|#1\rangle\langle#2|}

\begin{document}

\title{Single scatterings in single artificial atoms: Quantum coherence and entanglement}

\author{Ulrich Hohenester}\email{ulrich.hohenester@uni-graz.at} 
\author{Claudia Sifel}

\affiliation{Institut f\"ur Theoretische Physik, 
Karl-Franzens-Universit\"at Graz, Universit\"atsplatz 5, 8010 Graz, Austria}

\author{Pekka Koskinen}
\affiliation{Department of Physics, University of Jyv\=askyl\=a, 40014 Jyv\=askyl\=a, Finland}

\date{August 18, 2003}

\begin{abstract}

We employ the quantum-jump approach to study single scatterings in single semiconductor quantum dots. Two prototypical situations are investigated. First, we analyze two-photon emissions from the cascade biexciton decay of a dot where the single-exciton states exhibit a fine-structure splitting. We show that this splitting results for appropriately chosen polarization filters in an oscillatory behavior of two-photon correlations, and carefully examine the proper theoretical description of the underlying scattering processes. Secondly, we analyze the decay of a single-electron charged exciton in a quantum dot embedded in a field effect structure. We show how the quantum properties of the charged exciton are transferred through tunneling and relaxation to the spin entanglement between electrons in the dot and contact, and identify the pertinent disentanglement mechanisms.

\end{abstract}


\maketitle

\section{Introduction}

Ultrafast semiconductor spectroscopy provides a unique laboratory for the investigation of coherence and scattering effects in solids:~\cite{shah:96,rossi:02} a short laser pulse creates electron-hole pairs propagating in phase with the exciting laser; this initial phase coherence is consecutively destroyed through mutual interactions among the photoexcited carriers and couplings to environmental degrees of freedom, e.g. phonons, thus leading to {\em decoherence},~\cite{omnes:92,zurek:03} which can be monitored again by means of ultrafast spectroscopy. Recent years have seen spectacular examples of such ultrafast semiconductor spectroscopy, e.g. the observation of phonon quantum beats~\cite{banyai:95,wehner:98,rossi:02} or the buildup of screening,~\cite{huber:01} and have revealed dephasing and relaxation times of the order of femto- and picoseconds for carrier-carrier and carrier-phonon interactions, respectively.

In semiconductor quantum dots,~\cite{hawrylak:98,bimberg:98} sometimes referred to as {\em artificial atoms},\/ the strong confinement in all three spatial directions results in a substantial suppression of decoherence as compared to higher-dimensional semiconductors, i.e., bulk, quantum-wells, or quantum wires. On the one hand, Coulomb interactions among carriers captured in the dot do not result in scattering but only give rise to energy renormalizations of the electron-hole few-particle states.~\cite{landin:98,dekel:98,bayer:00a,warburton:00,hartmann.prl:00,findeis.prb:01} On the other hand, at low temperatures phonon-mediated dephasing of the electron-hole states of lowest energy is of only minor importance,~\cite{borri:01,krummheuer:02} and can become even negligible in comparison to spontaneous emission of photons which occurs on a nanosecond time scale.~\cite{bonadeo:98a,bonadeo:98b} These remarkable features render quantum dots ideal candidates for the solid-state implementation of quantum engineering,~\cite{bouwmeester:00,bennett:00} e.g., for the purpose of quantum computation, which has led to considerable research activities in recent years.~\cite{barenco:95,imamoglu:99,troiani.prb:00,biolatti:00,piermarocchi:02,troiani.prl:03,pazy:03,calarco:03,li:03}

Observation of optical coherence effects in ensembles of dots is usually spoiled by inhomogeneous line broadening due to dot size fluctuations, with typical broadenings comparable to the level splittings themselves. To overcome this problem, within the last couple of years a number of experimental techniques were developed to allow the observation of single quantum dots, hereby establishing the rapidly growing field of {\em single-dot spectroscopy}.~\cite{zrenner:00} Such single-dot measurements are not only indispensable for the observation of quantum coherence and the implementation of quantum engineering, but also open the possibility to study decoherence and relaxation in {\em single}\/ quantum systems. The theoretical analysis of single quantum system, however, is more cumbersome as compared to ensembles of corresponding systems. In the latter case, one usually employs the framework of quantum transport theory or quantum kinetics~\cite{haug:96,shah:96,rossi:02} which describes the time evolution of macroscopic quantities, e.g., the interband polarization, directly accessible to experiment; here, the detailed evolution of one specific subsystem is completely irrelevant, which allows to submit the problem to the laws of statistical physics. On the other hand, for a single system one has to be more specific about the occurrence of single scatterings and their monitoring through a measurement apparatus. The question of how to theoretically describe such problems first arose almost two decades ago when it became possible to store single ions in a Paul trap and to continuously monitor their resonance fluorescence, and led to the development of the celebrated {\em quantum-jump approach}.~\cite{dum:92,dalibard:92,plenio:98} This approach combines the usual master-equation approach with the rules of demolition quantum measurements,~\cite{hegerfeldt:93,plenio:98} and provides a flexible tool for the description of single-system dynamics subject to continuous monitoring.

In this paper we employ the quantum-jump approach to two representative examples for the observation of single scatterings in single semiconductor quantum dots. First, we study the cascade decay of a biexciton, which has recently attracted considerable interest:~\cite{benson:00,moreau:01,gywat:02,stace:03} in a first step, the biexciton decays through emission of a photon to the spin-degenerate exciton states; this final-state ambiguity leads to an {\em entanglement}\/ of photon and exciton; when subsequently the exciton decays, the photon-exciton entanglement is transferred to a two-photon entanglement.~\cite{gywat:02,benson:00} As shown below, for dots with an exciton fine-structure splitting~\cite{bayer:02} the two-photon correlation function of this cascade decay exhibits an oscillatory behavior, provided that the photons are detected in a properly chosen polarization basis (see Ref.~\onlinecite{flissikowski:01} for a similar experiment). Our analysis carefully examines the photon emission processes, and improves upon the simple-minded framework of Fermi's golden rule to provide a proper description of the problem under concern. Our second example is based on an experiment recently performed by Zrenner {\em et al.}~\cite{zrenner:02} in which the authors used a quantum dot embedded in a field-effect structure to transform an optically excited exciton through tunneling into a photocurrent. Here, we study the decay of a charged exciton, i.e., of a Coulomb-renormalized complex consisting of two electrons and a hole.~\cite{warburton:00,hartmann.prl:00,findeis.prb:01} Through the tunneling process one electron with either spin-up or spin-down orientation is transferred from the dot to the reservoir; since the electron in the dot has an opposite spin orientation, the spins of the reservoir- and dot-electron become entangled.~\cite{sifel.apl:03} In contrast to the optical biexciton decay additional difficulties arise because the tunnel-generated electron and hole do not propagate freely, as photons would in the corresponding scheme, but are subject to interactions in the contact. We develop a prototypical, though simplified, description scheme of the scattering channels in the reservoir, and demonstrate that this spin entanglement is robust against spin-unselective scatterings and is only deteriorated by spin-selective scatterings.

We have organized our paper as follows. In Sec.~\ref{sec:two.photon} we discuss the monitored biexciton cascade decay in a single quantum dot. We specify our model system, provide a short discussion of the master-equation and quantum-jump approach, and finally show how to obtain the system's time evolution through unraveling of the master equation. The charged-exciton decay of a dot embedded in a field-effect structure is discussed in Sec.~\ref{sec:spin.entanglement}. We develop a theory accounting for the complete cascade process of the buildup of coherence through tunneling, the swapping of quantum coherence to the spin entanglement through dephasing and relaxation, and the process of disentanglement through spin-selective scatterings. Finally, in Sec.~\ref{sec:conclusion} we draw some conclusions.

\section{Two-photon correlations}\label{sec:two.photon}

\subsection{Quantum dot level scheme}

\subsubsection{Linear basis}

\begin{figure}
\centerline{\includegraphics[width=0.5\columnwidth,bb=95 166 500 695]{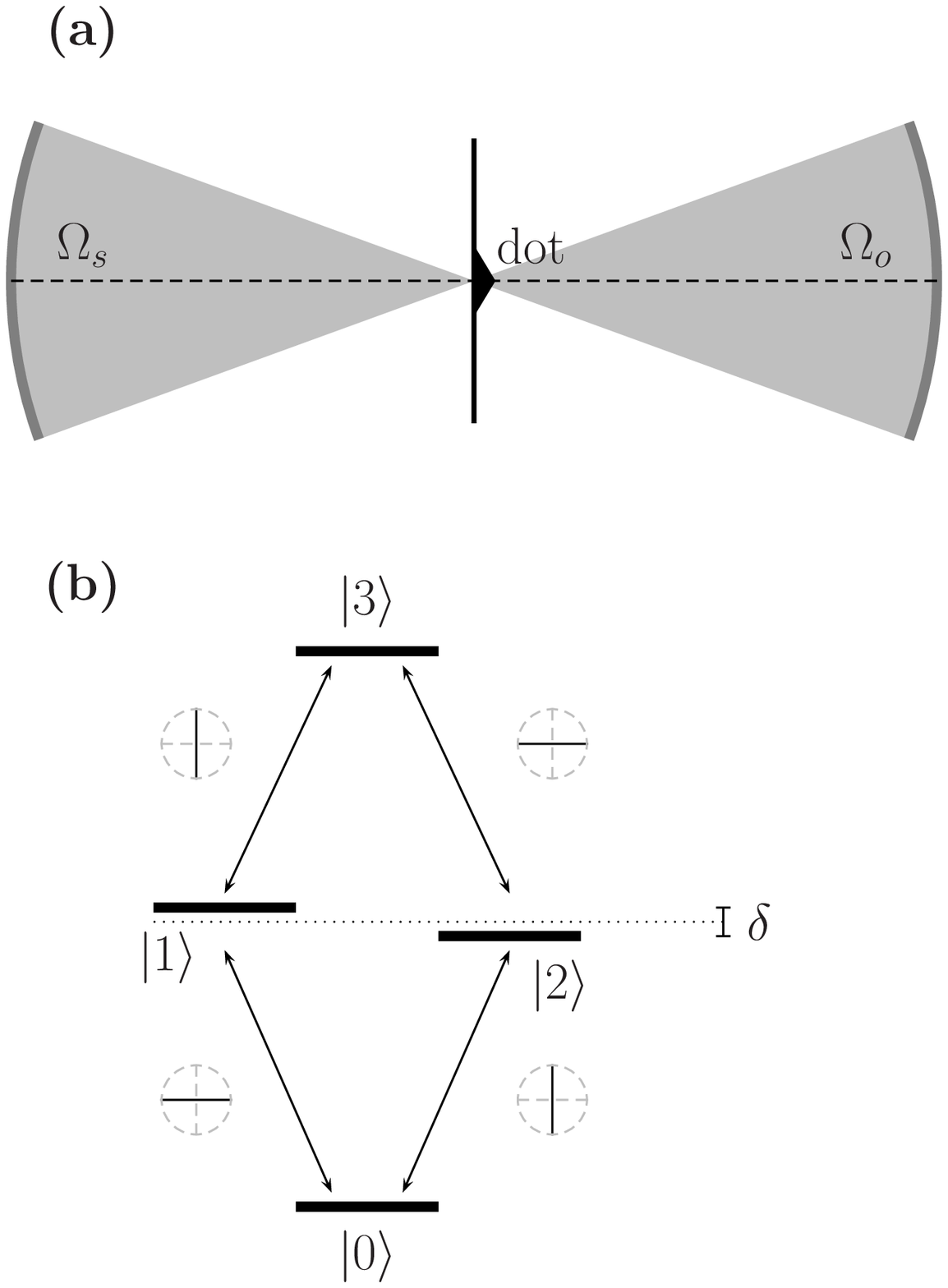}}
\caption{(a) Schematic picture of the proposed setup. Photons from the biexciton cascade decay are monitored by the photodetectors $\Omega_s$ and $\Omega_o$, where $\Omega_s$ ($\Omega_o$) is sensitive to photons from the 3--$s$ ($s$--0) decay. (b) Level scheme as described in the text: $0$ is the groundstate, $s=1,2$ the spin-degenerate single-exciton states, and $3$ the biexciton one. The arrows indicate the optically allowed transitions and the symbols indicate $x$- or $y$-polarization for photons propagating along $z$.}\label{fig:biexciton}
\end{figure}

We examine a prototypical level scheme depicted in Fig.~\ref{fig:biexciton}b, which consists of the groundstate $0$ (no electron-hole pairs present), the spin-degenerate single-exciton states $s=1,2$ of lowest energy, and the biexciton state $3$; other dot states, such as excited excitons or biexcitons, are energetically well separated and can be safely neglected. The single-exciton energies $\epsilon_s=E_o\pm\delta/2$ are assumed to be split by a small amount $\delta$ (typically a few tens of $\mu$eV) because of the {\em electron-hole exchange interaction},~\cite{bayer:02}\/ and the corresponding states to be polarized along $x$ and $y$. Finally, the biexciton energy $\epsilon_3=2E_o-\Delta$ is reduced by the {\em biexciton binding energy}\/ (typically a few meV) because of Coulomb correlations.~\cite{bayer:00a} Within this scheme, the Hamiltonian of the quantum-dot states reads:

\begin{equation}\label{eq:level.biexciton}
  H_o^d=\sum_{s=1,2}\epsilon_s\,\proj s s + \epsilon_3\,\proj 3 3\,,
\end{equation}

\noindent where energy zero is given by the groundstate. In what follows, we shall make use of the fact that the groundstate is optically connected to the biexciton state $3$ through two independent pathways, namely 0--1--3 and 0--2--3, which allows to establish a {\em quantum coherence}\/ between the single-exciton states directly observable in optical experiments. 

Because of the important role of states $s=1,2$ we shall find it convenient to introduce the Pauli matrices 
$\sigma_1=\proj 1 2 + \proj 2 1$,
$\sigma_2=-i\,(\proj 1 2 - \proj 2 1)$,
$\sigma_3=\proj 1 1 - \proj 2 2$,
which together with the unit matrix
$\openone=\proj 1 1 + \proj 2 2$
form a complete basis within the single-exciton subspace. The Hamiltonian of Eq.~(\ref{eq:level.biexciton}) can then be rewritten as

\begin{equation}\label{eq:level.biexciton.linear}
  H_o^d=E_o\,\openone+\frac \delta 2\,\sigma_3+\epsilon_3\,\proj 3 3\,.
\end{equation}

\subsubsection{Circular basis}

In the linear basis the Hamiltonian (\ref{eq:level.biexciton.linear}) is diagonal. Hence, a laser pulse propagating in $z$-direction with polarization along $x$ or $y$ allows to selectively excite one of the two single-exciton eigenstates. On the other hand, any other light polarization will induce a superposition state, e.g.,

\begin{eqnarray}\label{eq:level.biexciton.circular}
  |+\rangle &=& \phantom{i\,}(|1\rangle+i|2\rangle)/\sqrt 2\nonumber\\
  |-\rangle &=& i\,          (|1\rangle-i|2\rangle)/\sqrt 2
\end{eqnarray}

\noindent for circular polarization $\sigma_\pm$. Further below we shall find it convenient to switch between the linear basis $s=1,2$ and the circular one, Eq.~(\ref{eq:level.biexciton.circular}), through a unitary transformation $U=(\openone+i\sigma_1)/\sqrt 2$, where

\begin{equation}
  U\,(E_o\,\openone+\frac\delta 2\,\sigma_3)\,U^\dagger=
  E_o\,\openone+\frac\delta 2\,\sigma_2\,.
\end{equation}

\noindent In other words, the energy splitting $\delta$ translates in the circular basis to a coupling between states $\sigma=\pm$. Thus, if the system is initially prepared in, e.g., state $|+\rangle$ (which is not an eigenstate of the Hamiltonian), it will start to oscillate between $|\sigma=\pm\rangle$ where the oscillation period is given by $1/\delta$.

\subsection{Light-matter coupling}

We describe the light-matter coupling within the usual dipole and rotating-wave approximations~\cite{scully:97,hohenester:02}

\begin{equation}\label{eq:ham.op}
  H_{\rm op}=-\frac 1 2(
  \bm{\mathcal{E}}^{(-)}\,\bm{\mathcal{P}}+
  \bm{\mathcal{E}}^{(+)}\,\bm{\mathcal{P}}^\dagger)\,.
\end{equation}

\noindent Here, $\bm{\mathcal{E}}^{(\pm)}$ is the electric field of the light evolving with positive or negative frequency components,~\cite{mandel:95,scully:97}

\begin{equation}\label{eq:interband}
  \bm{\mathcal{P}}=\mu\sum_s\hat{\bm e}_s
  \left(\proj s 0 + \proj 3 {\bar s}\right)
\end{equation}

\noindent the {\em interband polarization}\/ of the quantum dot, $\mu$ the dipole moment, $\hat{\bm e}_s$ a unit vector along $x$ or $y$, and $\bar s$ denotes the state with polarization perpendicular to $s$.

\subsubsection{Coherent optics}

Suppose that the quantum dot initially in its groundstate is excited by a short laser pulse. We assume that $\bm{\mathcal{E}}^{(\pm)}$ is sufficiently intense to be treated classically, and that the spectral width of the pulse is sufficiently broad to excite both 0--$s$ transitions with comparable strength but narrow enough to inhibit transitions between $s$ and 3. Then, such a pulse with circular polarization $\sigma_+$ will induce a superposition $|\Psi\rangle=\alpha|0\rangle+\beta |+\rangle$,
where $\alpha$ and $\beta$ are determined by the laser parameters. Since $|+\rangle$ is not an eigenstate but a {\em coherent superposition}\/ of states $s=1,2$, as time goes on the system will oscillate between $|\pm\rangle$. 

This optically induced quantum coherence corresponds to an oscillating interband polarization which can be directly monitored by a second, time-delayed laser pulse (see, e.g., Refs.~\onlinecite{bonadeo:98b,lenihan:02} for the observation of such {\em coherence beats}). Let us assume that the frequency of the second pulse is tuned to the $s$--3 transitions and that the delay time is short as compared to the decoherence times. Then, the absorption of the second pulse will sensitively depend on the time delay between the two pulses, and will exhibit oscillations on a time scale of $1/\delta$.

\subsubsection{Incoherent optics}
\label{sec:incoherent}

In this paper we shall address the question whether it is possible to observe a similar effect by means of incoherent optics. We assume that initially the dot is in the biexciton state~\cite{remark:biexciton} and consider the setup depicted in Fig.~\ref{fig:biexciton}a, where the optical decay of photons along $\pm z$ is monitored by two photo detectors. For this geometry the usual optical selection rules (to be discussed below) apply. In addition, we assume that photo detector $\Omega_s$ ($\Omega_o$) is sensitive to photons from the 3--$\bar s$ ($s$--0) decay,~\cite{moreau:01} which can be achieved through spectral filtering (and restricts the time resolution to $\sim 1/\Delta\ll 1/\delta$). Suppose that at time $t$ a photon with polarization $\sigma_-$ is detected from the biexciton decay. Analogously to the coherent excitation-scenario described above, we might assume that after photon detection the system is prepared in state $|+\rangle$. Since this is not an eigenstate of the Hamiltonian, the system will oscillate between $|\pm\rangle$ which could be detected by monitoring the probability distribution for the second photon emission. Thus, time-resolved photon correlation measurements together with appropriate frequency and polarization filtering should provide a means for the measurement of quantum coherence in an incoherent optical experiment.

At this point we encounter a conceptual problem. If we would describe the photon emissions through Fermi's golden rule we would obtain scattering rates $\Gamma_{3\to \lambda}$ between the asymptotic states before, $3$, and after, $\lambda$, the scattering. Here, the photon correlations would strongly depend on the choice of the basis $\lambda$: in the circular basis, the detection of a $\sigma_-$ photon would correspond to an optical decay from 3 to $|+\rangle$ and would result in the aforementioned coherence beats. On the other hand, in the linear basis the detection of a $\sigma_-$ photon had to be associated with equal probability to either the 3--1 or 3--2 decay, and the state after photoemission would be a mixture {\em without any quantum coherence between $s=1,2$}\/ (since the nature of a scattering is intrinsically incoherent). Here is the problem: within the framework of Fermi's golden rule different predictions arise for different basis sets. This is a very unfortunate situation which calls for a more careful theoretical analysis. It is the purpose of the remainder of this section to reexamine the problem of photon correlation measurements from the biexciton decay and to provide a comprehensible answer.

\subsubsection{Photon-matter coupling}

We describe the coupling between the quantum-dot states and photons through Eq.~(\ref{eq:ham.op}), where the electric field of a photon with wavevector $\bm k$ and polarization $\sigma$

\begin{equation}\label{eq:efield.photon}
  \bm{\mathcal{E}}_{\bm k\sigma}^{(+)}=
  i\left(\frac{2\pi\omega_k}\kappa\right)^{\frac 1 2}
  \hat{\bm e}_{\bm k\sigma}\,a_{\bm k\sigma}
\end{equation}

\noindent is expressed in terms of the usual bosonic field operators $a_{\bm k\sigma}$.~\cite{mandel:95,scully:97,andreani:99} Here, $\omega_k=ck/n$ is the light frequency, $c$ the speed of light, $\kappa$ the semiconductor dielectric constant, $n=\sqrt\kappa$ its refractive index, and $\hat{\bm e}_{\bm k\sigma}$ the polarization vector (see also appendix \ref{sec:wigner-weisskopf}). With Eq.~(\ref{eq:efield.photon}) the light-matter coupling (\ref{eq:ham.op}) can be cast to

\begin{equation}\label{eq:ham.op.gamma}
  H_{\rm op}^\gamma=
  i\,\sum_{\bm k,\sigma s}\left(g_{\bm k,\sigma s}a_{\bm k\sigma}^\dagger
  \left(\proj 0 s+\proj {\bar s} 3\right)-\mbox{h.c.}\right)\,,
\end{equation}

\noindent with the coupling constant $g_{\bm k,\sigma s}=(2\pi\omega_k/\kappa)^{\frac 1 2}\mu\,(\hat{\bm e}_{\bm k\sigma}^*\,\hat{\bm e}_s)$. Additionally, in the following we will need the Hamiltonian for free photons $H_o^\gamma=\sum_{\bm k\sigma}\omega_k\, a_{\bm k\sigma}^\dagger a_{\bm k\sigma}$.

\subsection{Environment coupling}
\label{sec:environment}

Next, we address the problem of how to describe the time evolution of a single quantum system in contact with its environment, i.e., an {\em open system}.~\cite{walls:95,scully:97} Suppose that the Hamiltonian describing the system coupled to its environment can be split into the parts $H_o+V$, where $H_o$ accounts for the uncoupled system and reservoir (here $H_o^d+H_o^\gamma$) and $V$ to the coupling (here $H_{\rm op}^\gamma$). Because of the incoherent nature of such system-environment couplings we have to adopt a {\em density-matrix description},\/ where the density operator $\rho$ fully characterizes the system: its diagonal elements provide information about state occupancies whereas the off-diagonal elements account for quantum coherence. In lowest-order time-dependent perturbation theory and assuming that at time zero system and environment are effectively decoupled, the time evolution of $\rho$ is governed by:~\cite{walls:95,scully:97,breuer:99}

\begin{equation}\label{eq:open.system}
  \dot\rho_t^I\cong -\int_0^t dt'\,[V^I(t),[V^I(t'),\rho_t]\,]\,,
\end{equation}

\noindent where the superscript $I$ denotes the interaction representation according to $H_o$. Two points are worth emphasizing: first, Eq.~(\ref{eq:open.system}) is of Markovian form, i.e., the time evolution of $\rho$ at time $t$ is completely determined by the density operator itself; as discussed in detail in Ref.~\onlinecite{breuer:99}, this approximation is appropriate for systems where higher-order contributions to Eq.~(\ref{eq:open.system}) are negligible, or in other words, where scatterings occur seldomly on a time scale given by $H_o$ (here $\sim 1/\delta$); secondly, the time integral in Eq.~(\ref{eq:open.system}) extends to the past of the system and thus explicitly accounts for the buildup of scatterings. Note that the limit $t\to\infty$ in Eq.~(\ref{eq:open.system}), which we will {\em not}\/ perform below, would recover the framework of Fermi's golden rule and would lead to the aforementioned final-state ambiguities. 

Below we will need the expression

\begin{eqnarray}\label{eq:hop.interaction}
  &&\int_0^t dt'\,H_{\rm op}^I(t') = i\sum_{k,\sigma s}
  \int_0^t dt'\,(\nonumber\\
  &&\times g_{\bm k,\sigma s}a_{\bm k\sigma}^\dagger
  \,[\,\proj 0 s \,\phantom{{}^-}e^{i(\omega_k-\epsilon_s)t'}+
  \proj {\bar s} 3 \phantom{{}^-}\,
  e^{i(\omega_k+\epsilon_{\bar s}-\epsilon_3)t'}]\nonumber\\
  && 
  - g_{\bm k,\sigma s}^* a_{\bm k\sigma}
  \,[\,\proj s 0 \, e^{-i(\omega_k-\epsilon_s)t'}+
  \proj 3 {\bar s}\, e^{-i(\omega_k+\epsilon_{\bar s}-\epsilon_3)t'}]\,
  )\,,\nonumber\\ 
\end{eqnarray}

\noindent which accounts for the buildup of scatterings. The time integrals of the exponentials can be performed analytically,

\begin{equation}
  \int_0^t dt'\,e^{i\Omega t'}=\frac{e^{i\Omega t}-1}{i\Omega}\cong
  e^{i\Omega t}\gamma(\Omega,t)\,,
\end{equation}

\noindent with $\gamma(\Omega,t)=\sin\Omega t/\Omega$, and we have neglected in the last expression on the right-hand side terms which only contribute to an energy renormalization but not to dephasing and relaxation.~\cite{walls:95} $\gamma(\Omega,t)$ is a symmetric function of $\Omega$ which in the limit $t\to\infty$ gives Dirac's delta function. Finally, from Eq.~(\ref{eq:hop.interaction}) and the definition for $\gamma(\Omega,t)$ we obtain

\begin{eqnarray}\label{eq:op.decay}
  &&e^{iH_ot}\;\int_0^t dt'\,H_{\rm op}^I(t')\;e^{-iH_ot} = 
  i\sum_{k,\sigma s}
  \int_0^t dt'\,(\nonumber\\
  &&\times g_{\bm k,\sigma s}a_{\bm k\sigma}^\dagger
  \,[\,\proj 0 s \,\gamma(\omega_k-\epsilon_s,t)+
  \proj {\bar s} 3 \,
  \gamma(\omega_k+\epsilon_{\bar s}-\epsilon_3,t)]\nonumber\\
  &&
  - g_{\bm k,\sigma s}^* a_{\bm k\sigma}
  \,[\,\proj s 0 \, \gamma(\omega_k-\epsilon_s,t)+
  \proj 3 {\bar s}\, \gamma(\omega_k+\epsilon_{\bar s}-\epsilon_3,t)]
  \,)\,.\nonumber\\
\end{eqnarray}

\noindent This expression provides the starting point for our following discussion.

\subsubsection{Unmonitored decay}

We shall now be more specific about the question what happens to the emitted photon. Quite generally, the photon can propagate into a direction not covered by the photo detectors, in which case it will not be observed, or it can be measured by one of the two photo detectors, in which case we will acquire additional information about the state of the system. 

Let us first discuss the case of the {\em unmonitored decay},\/ i.e., photon emission into a segment not covered by the photo detector. For simplicity, we only consider the biexciton decay monitored by photo detector $\Omega_s$, where analogous conclusions hold for the exciton decay monitored by $\Omega_o$. In the spirit of the usual framework for the description of open-system dynamics,~\cite{walls:95,scully:97} we shall trace out the reservoir degrees of freedom. This procedure implies that we keep ignorant about the state of the reservoir.~\cite{zurek:03} Denoting with $\gamma\not\in\Omega_s$ the photon modes not covered by the photo detector, relaxation and dephasing in the dot due to unmonitored decay are described by the expression

\begin{equation}\label{eq:unmonitored}
  \dot\rho_t^{d,I}\cong
  -\mbox{tr}_{\gamma\not\in\Omega_s}\int_0^t dt'\,
  [V^I(t),[V^I(t'),\rho_t]\,]\,,
\end{equation}

\noindent where $\rho^d$ denotes the {\em reduced density operator}\/ of the quantum dot. Eqs.~(\ref{eq:op.decay},\ref{eq:unmonitored}) give us the prescription of how to compute the generalized scattering rates. Since each system--reservoir coupling $V$ comes with the bosonic field operators $a$ and $a^\dagger$, the resulting expressions contain expectation values of $aa$, $a^\dagger a^\dagger$, $a a^\dagger$, and $a^\dagger a$. We shall treat the reservoir (photons) as a thermal bath at zero temperature, i.e., we assume the usual correlation functions~\cite{walls:95,scully:97}

\begin{subequations}\label{eq:correlation.reservoir}
\begin{eqnarray}
  &&\mbox{tr}_\gamma\rho a_{\bm k\sigma}^\dagger\, a_{\bm k'\sigma'}\cong
    \mbox{tr}_\gamma\rho a_{\bm k\sigma}        \, a_{\bm k'\sigma'}\cong 0\\
  &&\mbox{tr}_\gamma\rho a_{\bm k\sigma} \, a_{\bm k'\sigma'}^\dagger\cong
    \delta_{\bm k\bm k'}\delta_{\sigma\sigma'}\,.
\end{eqnarray}
\end{subequations}

\noindent With these correlations it becomes possible to decompose Eq.~(\ref{eq:unmonitored}) into two parts, which, in the spirit of the Boltzmann equation, can be considered as generalized out- and in-scatterings. We start by evaluating the double commutator $[V,[V',\rho]]=VV'\rho+\rho V'V-V\rho V'-V'\rho V$. When tracing over the photon degrees of freedom, we realize that the only non-vanishing terms are proportional to $aa^\dagger\rho$, $\rho a a^\dagger$, and $a^\dagger\rho a$. Here, the first two terms do not change the number of photons. Correspondingly, we shall ascribe the contributions $VV'\rho+\rho V'V$ to {\em generalized out-scatterings}\/ which lead to decoherence and relaxation in the dot. If only these terms would be considered in the time evolution of $\rho^d$, the trace of the density operator, which describes the probability of finding the system in any of the quantum-dot states, would decrease. This lack of norm conservation is restored by the remaining terms $V\rho V'+V'\rho V$ which describe the effects of photon emissions, and may thus be interpreted as {\em generalized in-scatterings}.\/ As will be shown in Sec. \ref{sec:unraveling}, this decomposition into out- and in-scatterings allows for an efficient solution scheme through an {\em unraveling of the master equation}.~\cite{walls:95}

\subsubsection{Monitored decay}

The case of a {\em monitored decay}\/ has to be treated with slightly more care. Quite generally, such monitoring requires to specify in more detail the quantum mechanical measurement process. To this end, in the following we shall employ the framework of the {\em quantum-jump approach}.~\cite{hegerfeldt:93,plenio:98,hohenester.ssc:01} We assume that the photon counting can be approximately described by a series of repeated gedanken measurements at times $t_n=n\,\Delta_t$. Here, $\Delta_t$ is assumed to be sufficiently short to allow for a time-resolved measurement of the system's dynamics, i.e., $\Delta_t\ll 1/\delta$. On the other hand, $\Delta_t$ should be sufficiently large to enable the buildup of complete scatterings, such that the measurement process itself does not inhibit or enhance scatterings ({\em quantum Zeno effect}).~\cite{hegerfeldt:93,plenio:98}

Suppose that at time $t$ the density operator $\rho_t$ corresponds to an uncoupled dot and reservoir. With the projector $\mathbb{P}_n^{\Omega_s}$ associated to the result of the measurement~\cite{hegerfeldt:93,omnes:92} (no photon count, $n=0$, or photon count, $n=1$), the density operator at a later time can be evaluated to~\cite{hegerfeldt:93,plenio:98,hohenester.ssc:01}

\begin{equation}\label{eq:monitored}
  \rho_t^d \longrightarrow -\mathbb{P}_n^{\Omega_s}
  \int_0^t dt'\,[V^I(t),[V^I(t'),\rho_t]\,]\,\mathbb{P}_n^{\Omega_s}\,.
\end{equation}

\noindent If within the time interval $\Delta_t$ no photon is detected from the biexciton decay, either no photon has been emitted (which corresponds to a projector-like measurement $\mathbb{P}_0^{\Omega_s}$ on the photon vacuum) or a photon has been emitted into a segment $\Omega\not\in\Omega_s$. On the other hand, photon detection in $\Delta_t$ corresponds to a projector-like measurement $\mathbb{P}_1^{\Omega_s}$ on the single-photon subspace. Through the measurement we know that the biexciton has decayed, and accordingly we have to modify the density operator to account for this increase of information.

We can again evaluate the double commutator $[V,[V',\rho]]$ in Eq.~(\ref{eq:monitored}) and split the resulting expressions into generalized out- and in-scatterings; however, within the context of monitoring these terms acquire a different interpretation: as regarding out-scatterings, a nonvanishing contribution only arises for projection on the photon vacuum, which becomes $\mbox{tr}_{\gamma\in\Omega_s}(VV'\rho+\rho V'V)$. This term alone is responsible for a decrease of $\mbox{tr}\rho_t^d$ associated to the diminishing probability of finding the system in the initial biexciton state.~\cite{plenio:98} To restore this lack of norm conservation, we have to additionally consider in-scatterings. The corresponding contribution, which is associated to photo detection, is non-zero only if projected on $\mathbb{P}_1^{\Omega_s}$, and becomes $\mbox{tr}_{\gamma\in\Omega_s}(V\rho V'+V'\rho V)$. This term gives us the prescription of how the density operator must change after photon detection.~\cite{hegerfeldt:93,plenio:98}

\subsection{Master equation}

With the results of the previous subsections we are now in the position to improve upon the simple-minded Fermi's golden rule framework. First, we note that for both unmonitored and monitored decay the resulting expressions are, e.g., of the form $\mbox{tr}_{\gamma\in\Omega'}VV'\rho$, where the photon modes extend over a given segment $\Omega'$. Let us consider one specific contribution

\begin{widetext}
\begin{eqnarray}\label{eq:trace.photon}
  {\sum_{\bm k\sigma}}' g_{\bm k,\sigma s}^* g_{\bm k,\sigma s'}
  \gamma(\omega_k-\epsilon_{s'},t)&=&
  {\sum_{\bm k\sigma}}'\frac {2\pi\omega_k}\kappa \mu^2
  (\hat{\bm e}_{\bm k\sigma}  \hat{\bm e}_s)
  (\hat{\bm e}_{\bm k\sigma}^*\hat{\bm e}_{s'})
  \gamma(\omega_k-\epsilon_{s'},t)\nonumber\\
  &=&\frac {2\pi\mu^2}\kappa (2\pi)^{-3}
  \int_{\Omega\in\Omega'}d\Omega\,
  (\hat{\bm e}_{\bm k\sigma}  \hat{\bm e}_s)
  (\hat{\bm e}_{\bm k\sigma}^*\hat{\bm e}_{s'})\,
  \int_0^\infty k^2dk\,\omega_k
  \gamma(\omega_k-\epsilon_{s'},t)\,,\nonumber\\
\end{eqnarray}
\end{widetext}

\noindent where we have explicitly evaluated the trace over the photon degrees of freedom and used the definition of Eq.~(\ref{eq:ham.op.gamma}) for the coupling constants $g_{\bm k,\sigma s}$. The quantity of primary interest is the integral over $k$, which can be rewritten as

\begin{equation}\label{eq:photon.spectral}
  \int_0^\infty k^2 dk\,\omega_k\gamma(\omega_k-\omega_o,t)=
  \frac {n^3}{c^3}\int_0^\infty\omega^3 d\omega\,\gamma(\omega-\omega_o,t)\,.
\end{equation}

\noindent Eq.~(\ref{eq:photon.spectral}) has a rather precise physical meaning: it accounts for the temporal buildup and decay of correlations between dot states and photons, and thus plays the role of a memory function for the reservoir. Assume that $t\omega_o\gg 1$ which can be easily fulfilled on the time scale of the system's dynamics (since $\omega_o\sim E_o\gg \delta$). Then $\gamma(\omega-\omega_o,t)$ is a function which is strongly peaked around $\omega-\omega_o$. We thus can approximately replace in Eq.~(\ref{eq:photon.spectral}) $\omega^3$ with $\omega_o^3$. Then,~\cite{mandel:95}

\begin{eqnarray}\label{eq:photon.spectral.delta}
  &&\frac {n^3}{c^3}\omega_o^3\int_0^\infty d\omega\;\gamma(\omega-\omega_o,t)
  \nonumber\\
  &&\quad\cong\left(\frac{n\omega_o}c\right)^3\int_{-\infty}^\infty d\omega\,
  \gamma(\omega-\omega_o,t)=\pi\left(\frac{n\omega_o}c\right)^3\,,\qquad
\end{eqnarray}

\noindent where in the last line we have extended the integral to $-\infty$. Eq.~(\ref{eq:photon.spectral.delta}) is the expression we were seeking for. It shows that generalized scattering rates can be computed without invoking the adiabatic approximation which would be problematic in view of the choice of asymptotic states. Rather it suffices for the problem of our concern to assume that the reservoir memory is sufficiently short-lived to be approximately treated as a delta correlation on the time scale of the system dynamics.

For the $\Omega$-integrals in Eq.~(\ref{eq:trace.photon}) we assume

\begin{eqnarray}
  \int_{\Omega\not\in\Omega_o}d\Omega\,
  (\hat{\bm e}_{\bm k\sigma}^*\hat{\bm e}_s )
  (\hat{\bm e}_{\bm k\sigma}  \hat{\bm e}_s')&\cong&
  (1-\beta)\frac{4\pi}3\delta_{ss'}\nonumber\\
  \int_{\Omega\in\Omega_o}d\Omega\,
  (\hat{\bm e}_{\bm k\sigma}^*\hat{\bm e}_s )
  (\hat{\bm e}_{\bm k\sigma}  \hat{\bm e}_s')&\cong&
  \beta\frac{4\pi}3
  (\hat{\bm e}_{\sigma}^*\hat{\bm e}_s)
  (\hat{\bm e}_{\sigma}  \hat{\bm e}_s')\,,
\end{eqnarray}

\noindent where two approximations have been employed: first, for the unmonitored decay we have extended the integral over the complete sphere and have accounted for the missing segment $\Omega_o$ through a reduction factor $1-\beta$. Accordingly, we have introduced a factor $\beta$ for the monitored segment $\Omega_o$, and have assumed that within $\Omega_o$ the polarization vectors are approximately $\hat{\bm e}_{\bm k\sigma}\cong\hat{\bm e}_\sigma$, where $\hat{\bm e}_\pm\propto \hat{\bm e}_1\pm i\hat{\bm e}_2$.

With these expressions we can cast the equation of motion for the reduced density-operator $\rho^d$ (henceforth we will omit the superscript $d$) to a master equation of Lindblad form,~\cite{walls:95,scully:97}

\begin{equation}\label{eq:lindblad}
  \dot\rho\cong -i (H_{\rm eff}\rho-H_{\rm eff}^\dagger\rho)
  +\sum_i L_i\rho L_i^\dagger\,,
\end{equation}

\noindent where the Lindblad operators $L_i$ for the various scattering channels are listed in table \ref{table:lindblad} and $H_{\rm eff}=H_o^d-\frac i 2 \sum_i L_i^\dagger L_i$ is a non-Hermitian Hamiltonian accounting for the dot states and out-scatterings, which can be simplified to $H_{\rm eff}=H_o^d-i\frac\Gamma 2(2\,\proj 3 3+\sum_s \proj s s)$.

\begin{table}[b]
\caption{
Lindblad operators for the unmonitored and monitored decay channels. We use $\kappa^2=(1-\beta)\Gamma$ and $\kappa'^2=\beta\Gamma$, with $\Gamma$ the Wigner-Weisskopf decay rate given in Appendix \ref{sec:wigner-weisskopf}.
}

\begin{ruledtabular}
\begin{tabular}{l|l|l}\label{table:lindblad}
decay channel & unmonitored decay & monitored decay \\
& & (polarization $\sigma$) \\
\tableline
biexciton &
$L_{3\to \bar s}=\kappa\,\proj {\bar s} 3$ &
$L_{3\to\bar\sigma}^{\Omega_s}=\kappa'\, \proj {\bar\sigma} 3$ \\
exciton &
$L_{s\to 0}=\kappa\,\proj 0 s$ &
$L_{\sigma\to 0}^{\Omega_o}=\kappa'\, \proj 0 \sigma$ \\
\end{tabular}
\end{ruledtabular}
\end{table}

\subsection{Unraveling of the master equation}
\label{sec:unraveling}

The decomposition of the master equation (\ref{eq:lindblad}) into out-scatterings, $H_{\rm eff}$, and in-scatterings, second term on the right-hand side, allows for a convenient solution scheme through unraveling:~\cite{walls:95,plenio:98} suppose that the system is initially in state $\rho^{(0)}$. Then, $\rho_t^{(0)}=U_{\rm eff}(t)\rho^{(0)} U_{\rm eff}^\dagger(t)$ accounts for the time evolution subject to the condition that no scattering has occurred, with $U_{\rm eff}(t)=\exp-iH_{\rm eff}t$. Apparently, this term alone does not preserve the norm $p^{(0)}(t)=\mbox{tr}\rho_t^{(0)}$ since it only describes the situation where the system remains unscattered. For unmonitored scatterings this lack of norm conservation can be restored by introducing a further term which accounts for in-scatterings. The probability $p^{(1)}$ for a scattering at time $t$ is proportional to the decrease of $p^{(0)}(t)$, i.e., $p^{(1)}(t)=d(1-p^{(0)}(t))/dt$, and we obtain

\begin{equation}\label{eq:unraveling}
  \rho^{(t)}\cong p^{(0)}(t)\rho_t^{(0)}+\int_0^t dt'\, 
  p^{(1)}(t') U(t,t')\rho_{t'}^{(1)} U(t',t)\,.
\end{equation}

\noindent Here, $\rho^{(1)}=\sum_i L_i\rho^{(0)}L_i^\dagger/\mbox{tr}(.)$ is the density operator after scattering and the denominator ensures $\mbox{tr}\rho=1$. Because of the incoherent nature of scatterings this term becomes a {\em mixture}\/ of all possible final states. Finally, $U(t,t')$ describes the ensuing system's propagation, which may again be solved through unraveling (see, e.g., Sec.~\ref{sec:spin.entanglement}). 

It is important to realize that the mixture in Eq.~(\ref{eq:unraveling}) is due to our ignorance about the state of the system and reservoir. As the scattering can  occur at any time with probability $p^{(1)}(t)$, we have to sum in Eq.~(\ref{eq:unraveling}) over all possible histories. A different situation arises if we acquire additional information about the state of the system by monitoring the reservoir, e.g., through photo detection. Here, the detection of a photon at time $t$ allows to reduce the density operator from a mixture to a single term $\rho\to L_i\rho L_i^\dagger/\mbox{tr}(.)$, where $i$ corresponds to the outcome of the measurement, i.e., according to the photon's energy and polarization. Note that the monitored scatterings are described identically to the unmonitored ones with the only exception of the aforementioned reduction of $\rho$ due to our increase of information. This similarity is not incidentally, and can be understood more deeply within the framework of pointer states.~\cite{zurek:03}

\subsection{Photon correlation statistics}

Next, we show how to compute the two-photon correlation function $g^{(2)}$ from the biexcitonic cascade decay. We assume that initially the system is in state $3$. $g^{(2)}$ is then obtained through a series of repeated measurements, where only those contribute in which photons from the same cascade decay are detected by $\Omega_s$ and $\Omega_o$. Accordingly, in the following we can disregard photon emissions into unmonitored segments.

\subsubsection{Biexciton decay}

With the initial density operator $\rho^{(0)}=\proj 3 3$ the time evolution conditional to no photon emission can be evaluated to $\rho_t^{(0)}=e^{-2\Gamma t}\proj 3 3$. When a photon with polarization $\sigma_-$ is detected at time $t$, the density operator accordingly has to change to $\proj + +$, i.e. a {\em wavefunction purification}~\cite{bouwmeester:00} due to the measurement occurs.

\subsubsection{Exciton decay}

Since $|+\rangle$ is not an eigenstate of $H_o^d$ the system will start to oscillate between $|\pm\rangle$. Again we can unravel the corresponding master equation. It turns out to be convenient to use the Pauli-matrix representation of the circular basis, Eq.~(\ref{eq:level.biexciton.circular}), within which the effective Hamiltonian reads $H_{\rm eff}=(E_o-i\frac\Gamma 2)\openone+\frac\delta 2\sigma_2$. Hence,

\begin{equation}
  e^{-iH_{\rm eff}t}=e^{-i(E_o-i\frac\Gamma 2)t}
  (\cos\frac{\delta t} 2\,\openone-i\sin\frac{\delta t} 2\,\sigma_2
  )
\end{equation}

\noindent accounts for the conditional time evolution of no photon emission. The corresponding density operator is of the form $\rho_t^{(0)}=\proj {\Psi(t)}{\Psi(t)}$, where

\begin{equation}\label{eq:psi.biexciton}
  |\Psi(t)\rangle=e^{-iH_{\rm eff}t}|+\rangle=e^{-i(E_o-i\frac \Gamma 2)t}
  (\cos\frac{\delta t} 2\,|+\rangle\,+\,\sin\frac{\delta t} 2\,|-\rangle)
\end{equation}

\noindent is a state whose norm decreases with time, $\|\,|\Psi(t)\rangle\,\|^2=e^{-\Gamma t}$. From Eq.~(\ref{eq:psi.biexciton}) we can draw two conclusions: first, the probability distribution to detect a second photon with any polarization exhibits a mono-exponential decay without any coherence beats; on the other hand, the probability distribution for the detection of a photon with well-defined polarization $\sigma_+$ is $|\langle\sigma_+|\Psi(t)\rangle|^2=e^{-\Gamma t}\cos^2\frac{\delta t} 2$, which indeed exhibits the characteristic quantum-coherence beats. Thus, quantum-coherence phenomena are indeed observable in appropriately designed incoherent optical experiments.

\subsection{Photon entanglement}

We are now in the position to analyze the shortcoming of Fermi's golden rule in describing the problem of our present concern. Let us first discuss the biexciton decay. From Fig.~\ref{fig:biexciton}b we observe that for this decay the dot-level scheme effectively reduces to a $\Lambda$-type scheme, with $3$ the upper state and $s=1,2$ the lower, approximately degenerate ones. Because of the final-state ambiguity within the photon emission process no `decision is taken' regarding the final state but rather a quantum-mechanical superposition is formed: the underlying state corresponds to an {\em entanglement}\/ between the emitted photon and the exciton. Importantly, for a reservoir with sufficiently short-lived memory, a good approximation for the system under concern, photon and exciton become decoupled on a short timescale, thus allowing for the master-equation approach employed.

When the exciton decays through emission of a second photon, the photon-exciton entanglement is transferred to the two photons. Provided that the photons propagate along directions where the optical selection rules apply, i.e., sectors $\Omega_s$ and $\Omega_o$, the two photons are in an entangled state; indeed, in Refs.~\onlinecite{benson:00,gywat:02,stace:03} the authors proposed such biexciton decay as an entangled two-photon source. Finally, since the quantum properties of the initial photon-exciton system are directly transferred to the two photons it is not necessary to detect the first photon before the emission of the second one,~\cite{remark:ns} as one might naively assume from our discussion of the quantum-jump approach. Accordingly, the proposed scheme would also work for photo detectors placed far away from the dot.

\section{Spin entanglement}
\label{sec:spin.entanglement}

\begin{figure}
\includegraphics[width=0.65\columnwidth,bb=115 320 445 645]{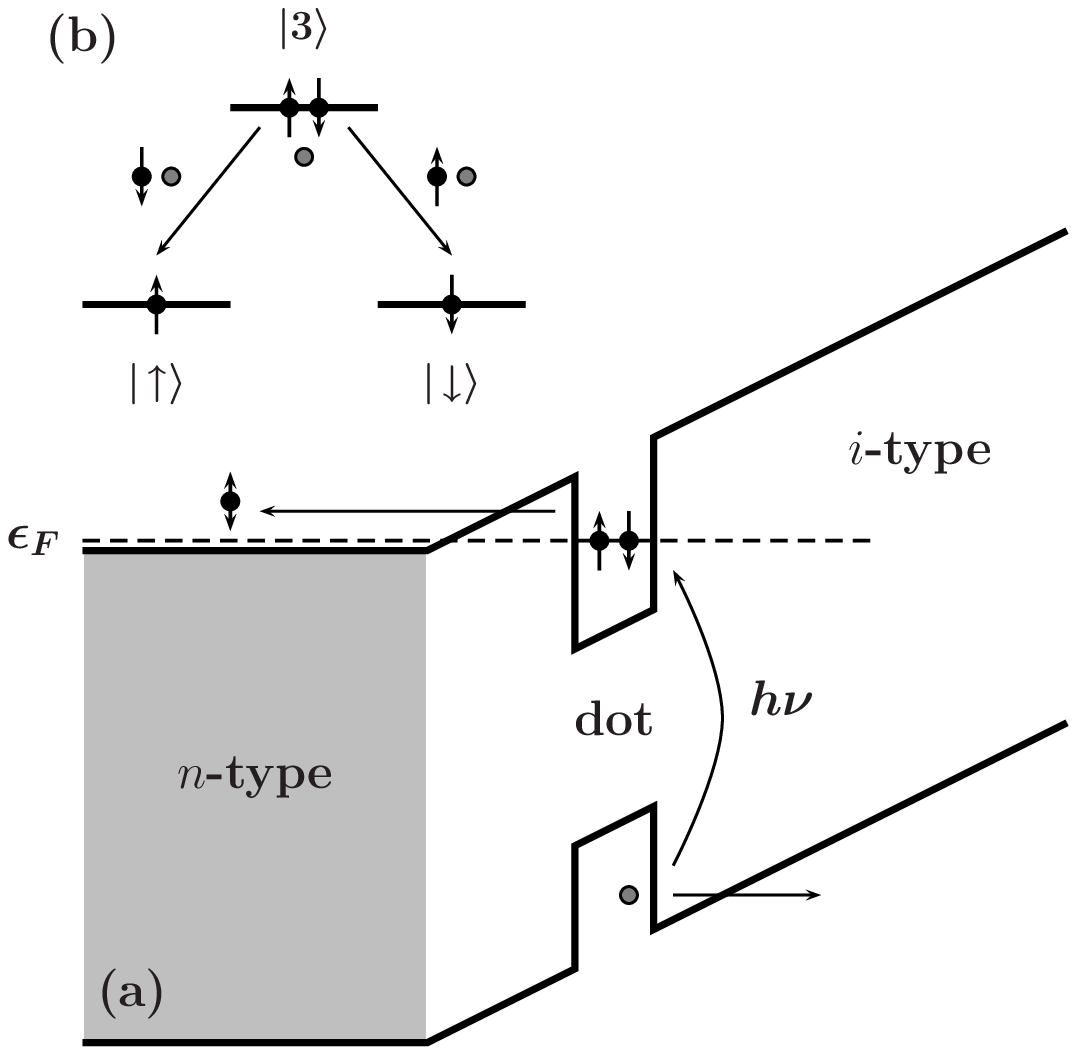}
\caption{(a) Schematic band diagram of the proposed structure. (b)
Level scheme of the spin-degenerate electron states
$\sigma$ and the charged-exciton state $3$ in the dot.}
\label{fig:entangler}
\end{figure}

In our second example we discuss the optically triggered spin entanglement of electrons. Consider the system depicted in Fig.~\ref{fig:entangler}a which consists of a single quantum dot embedded in a field-effect structure. We assume that the dot is initially populated by a single surplus electron, which can be achieved by applying an external bias voltage such that an electron is transferred from an nearby $n$-type reservoir to the dot;~\cite{warburton:00,findeis.prb:01} further charging is prohibited because of the Coulomb blockade. Optical excitation of this structure then results in the excitation of a {\em charged exciton},\/ i.e., a complex consisting of two electrons and a single hole;~\cite{warburton:00,hartmann.prl:00,findeis.prb:01} appropriate tuning of light polarization and frequency allows to selectively excite the charged-exciton groundstate, where the two electrons have opposite spin orientations. Since within the field-effect structure the charged exciton is not a stable configuration, in a consequent step one electron and hole will tunnel out from the dot to the nearby contacts; here, the system can follow two pathways, where either the electron in the dot has spin-up and the one in the reservoir spin-down orientation or vice versa. Similarly to the scheme discussed in the previous section, this final-state ambiguity results in a {\em maximal spin entanglement}\/ between the electrons in the dot and reservoir. 

However, as compared to the biexciton decay additional difficulties arise because the tunnel-generated electron and hole do not propagate freely, as photons would in the corresponding scheme, but are subject to interactions in the contact. As will be shown in the following, the framework developed in the previous section can be easily extended to describe such propagation in a solid-state environment, and we will provide a comprehensible theory accounting for the complete cascade process of: the buildup of three-particle coherence through tunneling; the swapping of quantum coherence to spin entanglement through dephasing and relaxation in the reservoirs; and finally the process of disentanglement through spin-selective scatterings. Since the main emphasis of our work is on the identification of the basic schemes underlying the buildup and decay of entanglement, we will rely on a simplified description scheme of environment interactions which will allow us to derive analytic expressions throughout.

\subsection{Level scheme}

Our model system comprises the dot, $H_o^d$, and the reservoir degrees of freedom for electrons and holes, $H_o^R$. The dot is characterized by the spin-degenerate single-electron states $\sigma$ with energies $\epsilon_\sigma$, and the charged-exciton state $3$ with energy $\epsilon_3$, i.e.,

\begin{equation}
  H_o^d=\sum_{\sigma=1,2}\epsilon_\sigma \proj \sigma \sigma + \epsilon_3 \proj 3 3 \,.
\end{equation}

\noindent The electron states in the reservoir are labeled by the continuum of quantum numbers $k$ and the spin index $\sigma$, where the fermionic field operator $c_{k\sigma}^\dagger$ creates an electron with energy $\epsilon_{k\sigma}^e$. Correspondingly, $d_k^\dagger$ creates a hole in state $k$; note that the quantum states $k$ for electrons and holes are different. Then,

\begin{equation}
  H_o^R=\sum_{k\sigma}\epsilon_{k\sigma}^e\, c_{k\sigma}^\dagger c_{k\sigma}+
  \sum_k\epsilon_k^h\, d_k^\dagger d_k\,.
\end{equation}

\noindent In the following we shall not specify the details of the reservoir states $k$, but rather pursue a simplified description scheme which only grasps the essential features. To this end, we introduce the fermionic field operators 
$C_\sigma(\omega)=\sum_k c_{k\sigma}\,\delta(\omega-\epsilon_{k\sigma}^e)$ for electrons and $D(\omega)=\sum_k d_k\,\delta(\omega-\epsilon_k^h)$ for holes, which average over all quantum states $k$ with the same energy $\omega$. The following anticommutation relations apply,

\begin{eqnarray}
  && \{C_\sigma(\omega),C_{\sigma'}(\omega')\}=
     \{C_\sigma^\dagger(\omega),C_{\sigma'}^\dagger(\omega')\}=0\nonumber\\
  && \{C_\sigma(\omega),C_{\sigma'}^\dagger(\omega')\}=
     g_\sigma^e(\omega)\delta_{\sigma\sigma'}\delta(\omega-\omega')\,,
\end{eqnarray}

\noindent with $g_\sigma^e(\omega)=\sum_k \delta(\omega-\epsilon_{k\sigma}^e)$ the electron density of states. Analogous relations apply for holes.

\subsection{Tunneling}

Since within the field-effect structure the charged exciton is not a stable
configuration, in a consequent step one electron and hole will tunnel out from the dot to the nearby contacts. For simplicity, we assume combined electron-hole tunneling, although our results would not be significantly modified in case of separate electron and hole tunneling.~\cite{sifel.apl:03} Then, 

\begin{equation}\label{eq:tunneling.k}
  V_t=-\sum_{kk',\sigma}\left(t_{kk',\sigma}c_{k\sigma}^\dagger d_{k'}^\dagger 
  \proj {\bar \sigma} 3 + \mbox{h.c.} \right)\,,
\end{equation}

\noindent where $t_{kk',\sigma}$ is the matrix element associated to the transition from the charged exciton state $3$ to $\bar\sigma$ through emission of one electron-hole pair. In Eq.~(\ref{eq:tunneling.k}) we have assumed that the total electron spin is a good quantum number which is conserved in the tunneling process. We furthermore assume that $t_{kk',\sigma}$ only depends on the electron and hole energies $\epsilon_{k\sigma}^e$ and $\epsilon_{k'}^h$, respectively, i.e., $t_{kk',\sigma}=\hat t(\epsilon_{k\sigma}^e,\epsilon_{k'}^h)$. Hence, the Hamiltonian of Eq.~(\ref{eq:tunneling.k}) can be rewritten as

\begin{equation}
  V_t\cong -\sum_\sigma\int d\omega_e d\omega_h\left(
  \hat t(\omega_e,\omega_h) C_\sigma^\dagger(\omega_e)D^\dagger(\omega_h)
  \proj {\bar\sigma} 3 + \mbox{h.c.}\right)\,.
\end{equation}

\noindent Note that all this simplifying assumptions could be easily lifted, however, at the price of more cumbersome notation.

\subsubsection{Master equation}

Assume that the initial density operator is $\proj 3 3$.~\cite{remark:charged.exciton}For low temperatures and sufficiently high tunneling rates we can safely neglect phonon processes and radiative decay in the dot, and tunneling becomes the only relevant scattering channel. In our theoretical approach tunneling is described, analogously to the photon emission of Sec.~\ref{sec:environment}, within lowest-order time dependent perturbation theory; in addition, we perform the adiabatic approximation~\cite{rossi:02} and replace $\gamma(\Omega,t)$ by $\pi\delta(\Omega)$. Since the hole enters with a high excess energy into the contact it immediately suffers an inelastic scattering, which guarantees that tunneling is an irreversible process. This is taken into account by tracing over the hole degrees of freedom and neglecting terms $\mbox{tr}_h\rho D^\dagger(\omega')D(\omega)\cong 0$. Then, $\mbox{tr}_h \rho D(\omega)D^\dagger(\omega')\cong g^h(\omega)\delta(\omega-\omega')$. Within this framework we arrive after some straightforward calculation at the master equation

\begin{eqnarray}
  &&\dot\rho_t\cong -i(H_{\rm eff}\rho_t-\rho_t H_{\rm eff}^\dagger)
  +\sum_{\sigma\sigma'}\int d\omega d\omega'\nonumber\\
  &&\times\left(\hat T_\sigma(\omega,\omega')+
                \hat T_{\sigma'}(\omega',\omega)\right)
  C_\sigma^\dagger(\omega)\proj {\bar \sigma} 3 \rho_t \proj 3 {\bar\sigma'}
  C_\sigma(\omega')\,,\nonumber\\
\end{eqnarray}

\noindent which includes tunneling as the only scattering channel. Here, $\hat T_\sigma(\omega,\omega')=\int g^h(\bar\omega)d\bar\omega\, \hat t(\omega,\bar\omega)\hat t(\omega',\bar\omega)\pi\delta(\omega+\bar\omega+\epsilon_{\bar\sigma}-\epsilon_3)$ and we have assumed that $\hat t$ is real. Furthermore, $H_{\rm eff}=H_o-i\sum_\sigma\int d\omega d\omega'\,\hat T_\sigma(\omega,\omega')\proj 3 3 C_\sigma(\omega')C_\sigma^\dagger(\omega)$ is an effective Hamiltonian which accounts for the dot and reservoir states, $H_o=H_o^d+H_o^R$, and outscatterings.

Next, we compute the term $H_{\rm eff}\rho$. We assume that the electrons in the reservoir can be described as an electron gas at zero temperature with total wavefunction $\prod_{k\sigma}\theta(\epsilon_F-\epsilon_{k\sigma}^e) c_{k\sigma} |0\rangle$, where Heaviside's step function $\theta$ ensures that only states with an energy below the Fermi energy $\epsilon_F$ are populated. Hence,

\begin{eqnarray}\label{eq:CCrho}
  &&C_\sigma(\omega')C_\sigma^\dagger(\omega)\,\rho\nonumber\\
  &&\quad=\theta(\omega-\epsilon_F)
  \left( g_\sigma^e(\omega)\delta(\omega-\omega')-
        C_\sigma^\dagger(\omega)C_\sigma(\omega')\right)\,\rho\nonumber\\
  &&\quad\cong \theta(\omega-\epsilon_F)\delta(\omega-\omega')g_\sigma^e(\omega)
  \,\rho\,.
\end{eqnarray}

\noindent The term $C_\sigma^\dagger(\omega)C_\sigma(\omega')\rho$ in Eq.~(\ref{eq:CCrho}) accounts for a process where an electron is exchanged between reservoir and dot. In the following we will neglect such processes since they only contribute to an energy renormalization but not to scattering and dephasing. From Fig.~\ref{fig:entangler} we infer that the energy of the tunneled electron must be larger than $\epsilon_F$ and smaller than some cutoff energy $\omega_c$, which is determined by the kinematics of the tunneling process. To simplify the following calculation, we assume that within the energy window $\epsilon_F<\omega,\omega'\le \omega_c$ the quantities $\hat T_\sigma(\omega,\omega')\cong \hat T$ and $g_\sigma^e(\omega)\cong g_o^e$ are approximately constant. Then, $H_{\rm eff}\cong H_o-i\Gamma_t'\proj 3 3$ with 

\begin{equation}
  \int_{\epsilon_F}^{\omega_c}d\omega\, g_\sigma^e(\omega)\hat T\cong
  g_o^e\,\Delta\,\hat T=\frac {\Gamma_t'} 2\,,
\end{equation}

\noindent where $\Gamma_t'$ denotes the single-electron tunneling rate and $\Delta=\omega_c-\epsilon_F$.

\begin{table*}
\caption{Contributions to the density operator corresponding to the different steps of the cascade process, i.e., for the initial charged exciton, the system after tunneling, after dephasing, and after disentanglement. The last column shows the time evolution of the corresponding probabilities.}

\begin{ruledtabular}
\begin{tabular}{l|l|l}\label{table:entangler}
step & density operator & time evolution of probability \\ 
\tableline
before tunneling & 
$\proj 3 3$ &
$\dot p^{(0)}=-\Gamma_t p^{(0)}$ \\
after tunneling &
$(2g_o^e\Delta)^{-1}
  \sum_{\sigma\sigma'}\int_{\epsilon_F}^{\omega_c}d\omega d\omega'
  C_\sigma^\dagger(\omega)\proj {\bar\sigma}{\bar\sigma'}C_\sigma(\omega')$ &
$\dot p^{(1)}=\Gamma_t p^{(0)}-\Gamma_d p^{(1)}$ \\
after dephasing &
$(2g_o^e\Delta)^{-1}
  \sum_{\sigma\sigma'}\int_{\epsilon_F}^{\omega_c}d\omega d\omega'
  C_\sigma^\dagger(\omega)\proj {\bar\sigma}{\bar\sigma'}C_\sigma(\omega')
  \,\delta(\omega-\omega')$ &
$\dot p^{(2)}=\Gamma_d p^{(1)}-2\kappa^2\Gamma_s p^{(2)}$ \\
after disentanglement &
$(2g_o^e\Delta)^{-1}
  \sum_{\sigma\sigma'}\int_{\epsilon_F}^{\omega_c}d\omega d\omega'
  C_\sigma^\dagger(\omega)\proj {\bar\sigma}{\bar\sigma'}C_\sigma(\omega')
  \,\delta(\omega-\omega')\,\delta_{\sigma\sigma'}$ &
$\dot p^{(3)}=2\kappa^2\Gamma_s p^{(2)}$ \\

\end{tabular}
\end{ruledtabular}
\end{table*}

\subsubsection{Unraveling of the master equation}

Within this scheme we can solve the master equation subject to the initial density operator $\proj 3 3$ through unraveling,

\begin{equation}\label{eq:rho.tunneled}
  \rho_t\cong e^{-\Gamma t}\proj 3 3+\Gamma_t\int_0^t dt'\,
  e^{-\Gamma_t t'} U(t,t')\rho^{(1)}_{t'}U(t',t)\,,
\end{equation}

\noindent where $\Gamma_t=2\Gamma_t'$ is the charged-exciton tunneling rate and $U(t,t')$ the time evolution operator accounting for the propagation of the tunneled electron (see below). The density operator after tunneling is $\rho^{(1)}=\proj{\Psi_1}{\Psi_1}$, with

\begin{equation}\label{eq:psi.entangled}
  |\Psi_1\rangle = (2 g_o^e \Delta)^{-\frac 1 2}\sum_\sigma
  \int_{\epsilon_F}^{\omega_c} C_\sigma^\dagger(\omega)|\bar\sigma\rangle\,.
\end{equation}

\noindent Eq.~(\ref{eq:psi.entangled}) is an important and non-trivial result. First, it demonstrates that despite the incoherent nature of tunneling and hole relaxation the electron system can be described in terms of wavefunctions; 
note that the detection of the hole would even allow to purify this wavefunction,~\cite{bouwmeester:00} which might be of relevance when initially $\rho$ is not equal to $|3\rangle\langle 3|$. Secondly, a closer inspection of
Eq.~(\ref{eq:psi.entangled}) reveals that the spin part $C_+|-\rangle+C_-|+\rangle$ is a {\em maximally entangled state}\/ of the electrons in the dot and reservoir. We emphasize that this maximal entanglement is independent of the
spin basis, which guarantees that our scheme is not deteriorated by possible polarization anisotropies of the dot states, e.g., fine-structure splittings.

\subsection{Dephasing}

After tunneling the electron in the reservoir propagates in presence of scatterings, as described by $U(t,t')$ in Eq.~(\ref{eq:rho.tunneled}). Quite generally, we assume that the orbital degrees of the reservoir electron are subject to much stronger interaction channels, e.g., phonons, than the spin
degrees, as evidenced by the long measured spin lifetimes ($\sim$ns) in $n$-doped semiconductors.~\cite{kikkawa:98} For that reason, let us first consider an elastic electron scattering which does not depend on spin, i.e., Lindblad operators of the form

\begin{equation}
  L(\omega)=\Gamma_d^{\frac 1 2}\sum_\sigma
  \frac{C_\sigma^\dagger(\omega)C_\sigma(\omega)}{g_o^e}\,,
\end{equation}

\noindent with $\Gamma_d$ the scattering rate. Hence, the effective Hamiltonian reads $H_{\rm eff}=H_o-\frac i 2\int d\omega\, L^\dagger(\omega) L(\omega)$. With $\int d\omega\,L(\omega)\proj{\Psi_1}{\Psi_1}\L^\dagger(\omega)/\mbox{tr}(.)$ the density operator after scattering and $L(\omega)|\Psi_1\rangle=(2 g_o^e \Delta)^{-\frac 1 2}\sum_\sigma C_\sigma(\omega) |\bar\sigma\rangle$, the master equation accounting for elastic spin-unselective scatterings can be solved through unraveling. Suppose that the tunneling has occurred at time $t'$, corresponding to one of the possible histories in Eq.~(\ref{eq:rho.tunneled}). Then,

\begin{eqnarray}
  &&U(t,t')\rho^{(1)}_{t'}U(t',t)=
  e^{-\Gamma_d(t-t')} e^{-iH_ot} \proj {\Psi_1}{\Psi_1} e^{iH_ot}\nonumber\\
  &&\qquad+\Gamma_d\int_{t'}^t
  d\bar t\,e^{-\Gamma_d(\bar t-t')} 
  U(t,\bar t)\rho^{(2)}_{\bar t}U(\bar t,t)
\end{eqnarray}

\noindent is the corresponding conditional density operator at later time. Here,

\begin{equation}\label{eq:rho.entangled}
  \rho^{(2)}_t=(2 g_o^e \Delta)^{-1}\sum_{\sigma\sigma'}
  \int_{\epsilon_F}^{\omega_c} 
  d\omega\, C_\sigma^\dagger(\omega)\proj {\bar\sigma}{\bar\sigma'}
  C_{\sigma'}(\omega)
\end{equation}

\noindent is the density operator after the elastic scattering and $U(t,\bar t)$ accounts for the propagation of the scattered electron. In comparison to Eq.~(\ref{eq:psi.entangled}) the density operator of Eq.~(\ref{eq:rho.entangled}) is diagonal in $\omega$, i.e., the elastic scattering has led to a destruction of the phase coherence (dephasing). However, the spin part still shows the same degree of entanglement, where similar conclusions would apply for inelastic but spin-independent scatterings. Thus, the decay of an optically excited charged-exciton indeed generates a robust spin entanglement between the electrons in the dot and reservoir.

\subsection{Disentanglement}

We finally comment on the process of disentanglement. In fact, any scattering channel which couples with unequal strength to the spins or affects only one spin orientation is responsible for such entanglement decay. To simplify our following analysis, we introduce operators acting in the reservoir subspace

\begin{equation}
  (g_o^e\Delta)^{-1}\int_{\epsilon_F}^{\omega_c} d\omega\,
  C_\sigma^\dagger(\omega)C_{\sigma'}(\omega)=\mathcal{P}_{\sigma\sigma'}\,,
\end{equation}

\noindent which are averaged over the orbital degrees of freedom $\omega$. Without specifying the details of such spin-selective scattering, we introduce the generic Lindblad operator

\begin{equation}\label{eq:lindblad.disentangle}
  L=\Gamma_s^{\frac 1 2}\left(\,(1+\kappa)\mathcal{P}_{11}+
                                (1-\kappa)\mathcal{P}_{22}\,\right)\,,
\end{equation}

\noindent where $\Gamma_s$ is a scattering rate and $\kappa$ a factor determining the asymmetry of the coupling to spin-up and spin-down electrons. As briefly sketched in Appendix \ref{sec:two-level}, the corresponding master equation can be solved analytically and we arrive at

\begin{eqnarray}\label{eq:disentangled}
  &&U(t,t')\rho^{(2)}_{t'}U(t',t)
  =e^{-2\kappa^2\Gamma_s (t-t')}
  \sum_{\sigma\sigma'}\mathcal{P}_{\sigma\sigma'}
  \proj {\bar \sigma}{\bar\sigma'}\nonumber\\
  &&\qquad+(1-e^{-2\kappa^2\Gamma_s (t-t')})\sum_\sigma
  \mathcal{P}_{\sigma\sigma}\proj {\bar\sigma}{\bar\sigma}\,.
\end{eqnarray}

\noindent Here, the first term on the right-hand side accounts for the unscattered state, whose probability decays with time, and the second term for the disentangled state after scattering. Because of our assumption for the Lindblad operators, Eq.~(\ref{eq:lindblad.disentangle}), the density operator after disentanglement still exhibits a classical correlation where the spins of the electrons in the reservoir and dot are antiparallel; apparently, it would require additional spin-flip processes to completely decouple the two electrons. See also Ref.~\onlinecite{sifel.apl:03} for a discussion of how such spin entanglement could be measured experimentally and possible quantum information applications.

\subsection{Cascade decay}

\begin{figure}
\centerline{\includegraphics[width=0.8\columnwidth,bb=135 475 490 685]{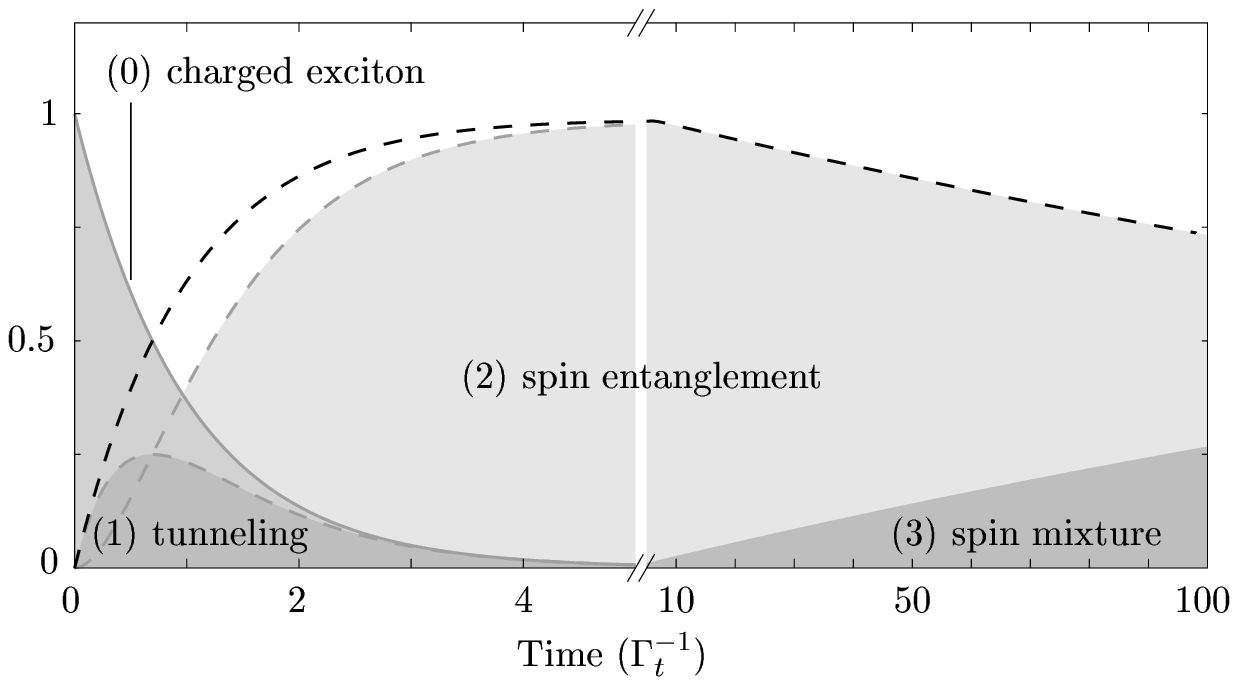}}
\caption{Schematic representation of the various contributions $p^{(i)}(t)$ of Eq. (\ref{eq:unraveling}) as discussed in the text. Times are measured in units of the inverse tunneling rate $\Gamma_t^{-1}$; we assume $\Gamma_d=2\,\Gamma_t$ for the spin-unselective and $2\kappa^2\Gamma_s=3\times 10^{-3}\,\Gamma_t$ for the spin-selective scatterings. The black dashed line shows the sum of $p^{(1)}(t)$ and $p^{(2)}(t)$ which corresponds to the degree of spin entanglement.
}\label{fig:entanglement}
\end{figure}

We finally list in table~\ref{table:entangler} the various contributions of the cascade decay to the density operator, i.e., corresponding to the step before tunneling, after tunneling, after dephasing, and after disentanglement. The probabilities for the corresponding contributions are determined by our unraveling procedure and can be obtained as follows: initially, the system is in the charged-exciton state, i.e., $p^{(0)}(0)=1$; because of tunneling $p^{(0)}$ decays mono-exponentially; correspondingly, the increase of the contribution after tunneling $p^{(1)}$ is proportional to the decrease of $p^{(0)}$, which again diminishes because of dephasing, and so forth. Within this scheme, we can easily obtain the time evolution of the various probabilities as shown in the last column of table~\ref{table:entangler}, with the initial conditions $p^{(0)}(0)=1$ and $p^{(i)}(0)=0$ for $i=1,2,3$. Fig.~\ref{fig:entanglement} sketches the various contributions for representative values of $\Gamma_d$ and $\Gamma_s$.

\section{Conclusion}
\label{sec:conclusion}

In conclusion, we have analyzed single scatterings in single semiconductor quantum dots for two representative examples. First, for the cascade decay of a biexciton it has been shown that an exciton fine-structure splitting results for appropriately chosen polarization filters in an oscillatory behavior of the two-photon correlations. In our second example we have investigated the tunneling decay of a charged exciton inside a field-effect structure into a photo current, and have shown that the spins of the electrons in the dot and reservoir become entangled. We have discussed that this entanglement is robust against dephasing and spin-unselective scatterings, and thus benefits from the long spin lifetimes in semiconductors.

Our theoretical approach has been based on a master-equation approach in Lindblad form in conjunction with the quantum-jump approach for the description of photon measurements. We have seen that the description scheme in terms of Lindblad operators, suitable for the environment couplings of our present concern, is superior over related quantum-transport descriptions based on the Markov and adiabatic approximations.~\cite{rossi:02} First, no assumption about the final states of a `scattering' has to be invoked, and it suffices to assume a sufficiently short-lived memory kernel of the reservoirs instead. Secondly, Lindblad operators associated to measurement, e.g., photon detection, allow for the application of the quantum-jump approach and provide a means to describe single-system dynamics. Finally, the distinction of out- and in-scatterings in the master-equation makes possible a flexible solution scheme through unraveling, which appears to be highly elegant in particular for the cascade processes discussed in this work. We expect our findings to be useful for the analysis of time-resolved single-system measurements and for the simulation of quantum-information applications.

\acknowledgments
Work supported in part by the {\em Fonds zur F\"orderung der wissenschaftlichen Forschung}\/ (FWF) under project No. P15752--N08.

\begin{appendix}

\section{Wigner-Weisskopf decay rate}
\label{sec:wigner-weisskopf}

In this appendix we show how to evaluate contributions of the form $2\pi\sum_{\bm k\sigma}g_{\bm k,\sigma s}^* g_{\bm k,\sigma s'}\,\delta(\omega_k-\omega_o)$. First, the polarization vector of a photon with wavevector $\bm k$ and polarization $\sigma$ can be expressed as~\cite{mandel:95}

\begin{eqnarray}
  \hat{\bm e}_{\bm k,+}&=&\frac 1 {\sqrt 2}
  (\cos\theta\cos\phi-i\sin\phi,
   \cos\theta\sin\phi+i\cos\phi,
  -\sin\theta)\nonumber\\
  \hat{\bm e}_{\bm k,-}&=&\frac i {\sqrt 2}
  (\cos\theta\cos\phi+i\sin\phi,
   \cos\theta\sin\phi-i\cos\phi,
   \phantom{-}\sin\theta)\nonumber\,.\\
\end{eqnarray}

\noindent The integral over the angular-dependent part can be evaluated to

\begin{eqnarray}
  &&\int_0^{2\pi}d\phi\int_0^\pi\sin\theta d\theta\,
  |\hat{\bm e}_{\bm k\sigma}^*\hat{\bm e}_s|^2=\frac{4\pi}3\nonumber\\
  &&\int_0^{2\pi}d\phi\int_0^\pi\sin\theta d\theta\,
  (\hat{\bm e}_{\bm k\sigma}^*\hat{\bm e}_1)
  (\hat{\bm e}_{\bm k\sigma}  \hat{\bm e}_2)=0\,,
\end{eqnarray}

\noindent i.e., only terms with $s=s'$ give a nonzero contribution. Finally, 

\begin{eqnarray}
  \Gamma &=& 2\pi\sum_{\bm k\sigma} |g_{\bm k\sigma}|^2 
  \delta(\omega_k-\omega_o)\nonumber\\
  &=&2\pi \frac{8\pi}3(2\pi)^{-3}\frac {2\pi\mu^2}{\kappa}
  \left(\frac{n\omega_o}c\right)^3
  =\frac{4n\mu^2\omega_o^3}{3c^3}\,,
\end{eqnarray}

\noindent is the Wigner-Weisskopf decay rate for a dipole radiator embedded in a medium with refractive index $n$.

\section{Master equation for two-level system}
\label{sec:two-level}

In this appendix we discuss the master equation of a generic two-level system. Since the Pauli matrices $\sigma_i$ together with the unit matrix $\openone$ form a complete basis, we can expand the density operator as

\begin{equation}
  \rho=u_0\openone+\bm u\cdot\bm\sigma\,,
\end{equation}

\noindent with the vector $\bm\sigma=(\sigma_1,\sigma_2,\sigma_3)$. From $\mbox{tr}\rho=1$ we find $u_0=\frac 1 2$ and $\rho=\rho^\dagger$ implies that $\bm u$ is real. Analogously, a generic Lindblad operator is of the form $L=a_0\openone+\bm a\cdot\bm\sigma$, where $a_0=a_0'+ia_0''$ and $\bm a=\bm a'+i\bm a''$ can be decomposed into real and imaginary parts. The master equation corresponding to this Lindblad operator is of the form

\begin{equation}
  \dot\rho=\frac 1 2\left([L\rho,L^\dagger]+[L,\rho L^\dagger]\right)\,,
\end{equation}

\noindent which can be easily evaluated by using the usual commutation relations for Pauli matrices. We arrive after some lengthy calculation at

\begin{eqnarray}
  \dot{\bm u}&=& 2\biggl( (\bm a'\times \bm a'')-
  (a_0'\bm a''-a_0''\bm a')\times\bm u\nonumber\\
  &&\qquad -|\bm a|^2\bm u+
  (\bm a'\bm u)\,\bm a'+(\bm a''\bm u)\,\bm a''\biggr)\,.
\end{eqnarray}

Then, for the Lindblad operator $L=\Gamma^{\frac 1 2}(\openone+\kappa\sigma_3)$ of Eq.~(\ref{eq:lindblad.disentangle}) for disentanglement we find $\dot u_{1,2}=-2\kappa^2\Gamma\, u_{1,2}$ and $\dot u_3=0$, which can be easily solved to arrive at the final result of Eq.~(\ref{eq:disentangled}).

\end{appendix}


\begin{thebibliography}{55}
\bibitem{remark:biexciton}
Such selective biexciton population could be achieved similar to quantum-dot based single-photon experiments:~\cite{michler:00,santori:01,moreau:01} first, a short laser pulse creates electron-hole pairs in continuum states in the vicinity of the quantum dot; carrier capture results in a multi-exciton complex within the dot, which decays through successive emission of photons; because of environment coupling, after each photon emission the system quickly relaxes to the electron-hole state of lowest energy. In what follows, it is sufficient to assume that during this cascade process the dot will eventually pass through state $3$.
\bibitem{remark:ns}
Note that within a typical exciton lifetime, i.e., nanoseconds, the distance travelled by a photon is of the order of 30 centimeters.
\bibitem{remark:charged.exciton}
Such selective charged-exciton occupation can be achieved as follows. Suppose that initially the spin direction of the doping electron is undetermined, i.e., the corresponding density operator is a mixture $\rho=\frac 1 2 \sum_\sigma\proj\sigma\sigma$. When at time $t=0$ the dot is subject to an unpolarized optical $\pi$-pulse~\cite{zrenner:02} the system will be excited in accordance to our assumption to state $\proj 3 3$.
\end{thebibliography}

\end{document}